\documentclass[article,amsart,fleqn,amsmath,floats,nofootinbib,aps,twocolumn,pra,floatfix,showpacs,superscriptaddress]{revtex4-2}
\usepackage{cmap}
\usepackage{wrapfig} 
\usepackage[pdftex]{graphicx}
\usepackage{pgfplotstable}
\usepackage[pdftex,unicode,colorlinks,bookmarks=false,citecolor=blue,linkcolor=blue,urlcolor=blue]{hyperref} 
\usepackage[T1,T2A]{fontenc} 
\usepackage{latexsym}
\usepackage{amsmath}
\usepackage{amssymb}
\usepackage{amsfonts}
\usepackage{color}
\usepackage{bm}
\usepackage{verbatim}
\usepackage[english]{babel}
\usepackage[utf8]{inputenc}
\usepackage{euscript}
\usepackage{subfigure}
\usepackage[normalem]{ulem}
\usepackage{listings}
\hypersetup{
colorlinks=true,
linkcolor=blue,
filecolor=magenta,
urlcolor=cyan,
}

\begin{document}
\title{Point-contact spectroscopy of electron-phonon interactions and anisotropy of superconducting energy gaps in $2H-NbS{{e}_{2}}$.}
\author{N.~L.~Bobrov, L.~V.~Tiutrina, \fbox{S.~B.~Feodos'ev}, V.~V.~Fisun}

\affiliation{B.I.~Verkin Institute for Low Temperature Physics and Engineering of the National Academy of Sciences of Ukraine\\ prosp. Nauki 47, 61103 Kharkiv, Ukraine}

\email{bobrov@ilt.kharkov.ua}

\begin{abstract}
{The superconducting state of $2H-NbS{{e}_{2}}$ was studied using point contacts for two main orientations of the contact axis relative to the principal crystallographic orientations: along the $c$-axis and in the $ab$ plane.
For contacts oriented in the $ab$ plane, dependencies proportional to the electron-phonon interaction function were obtained, which agree perfectly with the theoretical calculation. It was found that the features on the experimental curves associated with the electron-phonon interaction extend to energies twice the cutoff energy of the phonon spectrum and have an amplitude of the same order of magnitude throughout the entire energy range. It has been theoretically demonstrated that the features at energies exceeding the boundaries of the phonon density of states are caused by two-phonon scattering processes of non-equilibrium phonons on Andreev electrons. A criterion for the undisturbed nature of the crystal lattice in the region where gap features form has been found. Using $2H-NbS{{e}_{2}}-Cu$ heterocontacts in the superconducting state, a study of energy gaps was conducted for two orientations of the $NbS{e_2}$ crystal: along the $c$-axis and in the $ab$ plane. It was found that the superconducting energy gaps are anisotropic; specifically, three gaps were identified in the $ab$ plane, whereas only two were observed in the c direction.\\

 }.


\end{abstract}

\maketitle

\tableofcontents{}

\section{Introduction}\label{sec1}

The presence of a peculiar “probe”—electron duplication—is a necessary and sufficient condition for a direct-conductivity contact to be spectroscopic, allowing for the study of the spectra of electron interactions with quasiparticle excitations. The phenomenon of duplication, first reported in 1987 \cite{1}, means that in the current-carrying state, electrons (holes) in the vicinity of the contact separate into two groups, for which the energy difference between the occupied and free states at the Fermi surface is exactly equal to the applied voltage. The simplest example of such duplication for a point contact in the form of a hole for a spherical Fermi surface is shown in Fig.3 in \cite{2}.

For ballistic and diffusion point contacts between normal metals in a current-carrying state, a deviation from Ohm’s law is observed in the form of a negative current contribution due to processes of electron backscattering onto non-equilibrium phonons. Due to geometric limitations, such processes are effective only in the volume near the constriction. As the distance from the center of the contact increases, the probability of these processes decreases very rapidly. Since the energy relaxation length of electrons is significantly larger than the dimensions of the contact, most of the electron scattering processes on non-equilibrium phonons are not backscattering and do not contribute to the nonlinearity of the current-voltage curve ($I-V$ curve). The second derivative of the $I-V$ curve for such contacts is proportional to the electron-phonon interaction (EPI) function. The magnitude of the nonlinear deviation from Ohm’s law depends on the intensity of the EPI and amounts to several percent, but in any case does not exceed 10\%.

If one of the electrodes is in a superconducting state, then in addition to the nonlinearity of the $I-V$ curve caused by back-scattering processes, an additional nonlinearity appears in the phonon energy region, caused by the scattering of non-equilibrium phonons generated by electrons with excess energy equal to the voltage applied to the contact, on electrons undergoing Andreev reflection (Andreev electrons). Unlike backscattering processes, there are no geometric restrictions here. In principle, they are possible at any distance where non-equilibrium phonons and Andreev electrons coexist, i.e., up to the coherence length, at which the conversion of Andreev electrons into Cooper pairs occurs. Nevertheless, the theory \cite{3} considers only scattering processes occurring in the high-current-density region, where the concentrations of both components are high. For contacts in which the condition that the diameter is small compared to the coherence length is satisfied quite strictly, the magnitude of the nonlinear deviation of the contact resistance from Ohm’s law is several times smaller than the nonlinearity caused by backscattering processes. Therefore, the second derivative of the $I-V$ curve for such contacts in the phonon energy region changes only slightly when one of the electrodes transitions to the superconducting state.

It was theoretically predicted that the first derivative of the excess current (the difference between the $I-V$ curves of the superconducting and normal states) should be proportional to the EPI function \cite{3}. However, no such proportionality was observed either for the first derivative of the excess current or for the difference between the first derivatives of the $I-V$ curves of a point contact in the normal and superconducting states. It was not until nearly 30 years later that it was established \cite{4} that the sought-after proportionality is observed for the integral of the difference of the second derivatives of the $I-V$ curves after subtracting the superconducting background, which is not accounted for by the theory \cite{3}. The background curve must satisfy two basic requirements: first, the difference curve, after subtracting the background, must vanish outside the phonon spectrum; second, the curve obtained after subtracting the background must satisfy the geometric rule of sums—the sums of the areas above and below the $x$-axis must be equal. This requirement means that after integrating the difference curve with the background subtracted, we obtain a curve proportional to the EPI function with a zero background.

It is clear that many different background curves can satisfy these requirements; however, provided that the curves are smooth and lack any structure, all these variations do not lead to noticeable deviations in the relative intensity, shape, and position of the phonon features.

As previously noted, it follows from \cite{3} that the nonlinearity caused by the scattering of non-equilibrium phonons generated in the region of current concentration on Andreev electrons is significantly smaller than the nonlinearity caused by backscattering processes. The spectra for contacts in the normal and superconducting states differ only very slightly. For example, the spectra for tin-based contacts, with the exception of the presence of a superconducting background, agree perfectly with the predictions of the theory \cite{3}. At the same time, despite the similarity of such superconducting parameters as the gap or critical temperature, the second derivative for the tantalum-based contact undergoes radical changes during the transition to the superconducting state. The contribution of low-energy phonons increases sharply; instead of the predicted broadening of the phonon peaks, they become sharper and their positions are fixed near the maxima of the phonon density of states (see, for example, Fig. 2 in \cite{5}). This transformation is particularly pronounced for relatively low-resistance contacts (e.g., Figs. 8 and 9 in \cite{6}). The mechanism of nonlinearity formation in the current concentration region near the constriction in tantalum is the same as in tin; however, at the periphery, as one moves away from the current concentration region, the probability of scattering of non-equilibrium phonons on Andreev electrons turns out to be significantly higher than in tin.

The Andreev reflection is a quantum process in which Andreev electrons are converted into Cooper pairs within a volume of a size on the order of the coherence length. In the isotropic case, with spherical current spreading, in the region of space where an Andreev electron scatters off a non-equilibrium phonon, the probability density of the wave function of such an electron—all other conditions being equal—will be inversely proportional to the cube of the coherence length. For ballistic contacts based on tantalum, the volume of the sphere bounded by the coherence length is approximately 100 times smaller than that for tin (see Table 1 in \cite{6}). Thus, the rate at which the probability of non-equilibrium phonon scattering off Andreev electrons decreases as the current spreads out is here approximately $\sim 100$ times smaller. Since the scattering of a non-equilibrium phonon on an Andreev electron is not a point process but occurs in a volume of the order of the phonon’s wavelength, such processes are more likely to occur precisely for low-energy phonons with long wavelengths, which is what is observed in experiments. Furthermore, in scattering processes, phonons with low group velocities near the maxima of the phonon density of states make a larger contribution. They leave the contact region more slowly, and their concentration is higher, which leads to a sharpening of the phonon peaks. Therefore, the additional specific contribution of low-energy phonons and phonons with low group velocities to the negative add-on in the excess current is decisive, which explains the observed transformation of the phonon spectrum. The scattering probability of fast, high-energy phonons in this region of space increases only slightly (see Fig.2b in \cite{5}).

We emphasize that this type of selection of non-equilibrium phonons is observed only in ballistic contacts, in which fast, high-energy phonons leave the selection region freely without accumulating.

Since the low-frequency region of the spectrum undergoes the most radical transformation during the transition to the superconducting state, the resulting curve—after subtraction of the superconducting background—does not satisfy the geometric sum rule and requires correction (see, for example, Fig. 6 in \cite{6}). To do this, the curve obtained after subtracting the background was multiplied by a stepped empirical correction curve, the shape and magnitude of which were selected on a case-by-case basis.

If the nonlinearity of the $I-V$ curve, caused by backscattering processes, is comparable to the nonlinearity generated by the reabsorption of non-equilibrium phonons by Andreev electrons, and if it is difficult to bring the contact into a normal state, then problems arise with the recovery of the EPI function. Since backscattering processes and the reabsorption of non-equilibrium phonons in such contacts occur in geometrically different areas of space, it is possible to suppress the nonlinearity caused by backscattering processes without significantly affecting the reabsorption of non-equilibrium phonons by Andreev electrons. To achieve this, it is necessary to use inhomogeneous contacts. Here, an inhomogeneous contact refers to one in which the volume corresponding to the backscattering processes is in the diffusion regime, while the remaining region of the metal over the coherence length is undisturbed. The intensity of backscattering processes in the diffusion regime is reduced by a factor of approximately $\sim l_{i}/d$ compared to the ballistic regime, where $l_{i}$ is the mean free path, and $d$ is the contact diameter \cite{7}. The diffusion regime is achieved by a high concentration of defects/impurities in the narrowing region.

Corresponding experiments were conducted on niobium point contacts \cite{8}. In most cases, in the normal state, the second derivatives of the $I-V$-curves were curves without distinct features in the phonon energy region. When the contact transitioned to the superconducting state, the shape of the second derivative became similar to the difference between the second derivatives of the $I-V$ curves in the superconducting and normal states of tantalum point contacts, with some differences. Similarly, after subtracting the superconducting background to recover the EPI function, the obtained curves must be corrected so that they satisfy the aforementioned geometric sum rule. In \cite{8}, in contrast to the empirical selection of a correction curve, a quantitative approach was developed in which the curve was characterized by its magnitude—the effective area under the curve from zero to the boundary of the phonon spectrum (see Equation (1) in \cite{8}). As was shown, the correction result differed only slightly for three correction curves of different shapes, provided that their magnitude corresponded to the minimum value on the energy axis (Fig.4 from \cite{8}). It can be assumed that for each experimental curve requiring correction, a unique correction curve can be found whose magnitude is minimal, i.e., one that satisfies the principle of minimal necessary impact. This is undoubtedly of academic interest; however, for practical use, any of the curves proposed in \cite{8} can be applied. It is evident that the existence of a correction curve with the smallest possible magnitude also holds true for tantalum point contacts.

\section{The discovery of superconducting energy gaps in $S-c-N$ contacts}\label{sec2}

In $S-c-N$ point contacts, current flow is determined by the quantum process of Andreev reflection. Ideally, in the absence of electron scattering at the interface and at temperatures approaching zero, for voltages smaller than the superconducting energy gap, the contact conductivity doubles. The intermediate regime between barrier-free and tunneling current flow is described by the Blonder–Tinkham–Klapwjik (BTK) model \cite{9}. A generalized form of the BTK equation in the two-gap approximation, which allows for the finite lifetime of carriers by introducing the broadening parameter $\Gamma$, is given in \cite{10} (Equation 5). When there are several superconducting gaps on the Fermi surface, the total conductivity is a superposition of the conductivities from all these regions with corresponding weighting factors reflecting their partial contributions. The procedure for determining all the superconducting parameters of an experimental curve consists of selecting them such that the averaged root-mean-square deviation between the experimental and theoretical data points on the curves is minimized. Obviously, as the number of free fitting parameters increases, this task becomes easier. Therefore, one of the requirements for the calculation is to determine the necessary and sufficient number of such parameters.

To describe $S-c-N$ point contacts, modified versions of the BTK model were used in the two-gap (see Equation (5) in \cite{10}) and the three-gap approximation given below:

\begin{equation}
\label{eq1}
\begin{matrix}
\frac{{dV}}{{dI}} = \\ = \frac{{{S_F}}}{{\frac{{dI}}{{dV}}({\Delta _1},{\Gamma _1},{\rm{Z}}){K_1} + \frac{{dI}}{{dV}}({\Delta _2},{\Gamma _2},{\rm{Z}}){K_2} + \frac{{dI}}{{dV}}({\Delta _3},{\Gamma _3},{\rm{Z}}){K_3}}}
\end{matrix}
\end{equation}

In this equation, the dimensionless parameter $Z$ characterizes the value of the $\delta$-functional barrier at the boundary; the broadening parameter $\Gamma$ arises due to the finite carrier lifetime and has the same dimension as the superconducting energy gap. The parameter $Z$ in this model can vary from 0 to infinity; in fact, when $Z$ $\sim 10$, we have a tunneling contact. The broadening parameter $\Gamma$ leads to broadening and suppression of the curve intensities. The coefficient $K_{i}$ represents the contribution to conductivity from the Fermi surface region with a gap $\Delta_{i}$, where $K_{1}+K_{2}+K_{3} = 1$; $S_{F}$ is a scaling factor characterizing the ratio of the experimental curve’s intensity to the theoretical one and used to minimize their mutual root-mean-square deviation in shape. The experimental curves are approximated by this expression, from which the parameters $\Delta_{1,2,3}$, $\Gamma_{1,2,3}$, $Z$, $S_{F}$ and $K_{1,2,3}$ are extracted. Let us focus on the parameter $S_{F}$, which is usually overlooked in experimental studies. In addition to its shape, the theoretical curve—normalized to the normal state—has a unique amplitude or intensity for each set of the above-mentioned parameters. When the theoretical curve is fitted to the experimental curve by calculating the average root-mean-square deviation of the points on the curves, this parameter appears. It is not a fitting parameter, but rather an indicator of how closely the intensity (or amplitude) of the experimental curve matches the theoretical model. This factor is always present in the output array of calculation data when calculating the averaged root-mean-square deviation; it is only necessary to ensure that it is explicitly displayed along with the other calculation parameters. If it is less than 1, this indicates a deviation of the actual contact from the theoretical model; for example, not the entire volume is filled with a superconductor, or part of the superconductor has reduced superconducting properties, etc. As a rule, such deviations are insignificant and can be disregarded. If, however, it is greater than 1, this means that an attempt is being made to approximate the experimental curve with a blurred theoretical curve of lower intensity. In this case, it is justified to use a model with a larger number of gaps and less blurring.

\begin{figure}[]
\includegraphics[width=8.5cm,angle=0]{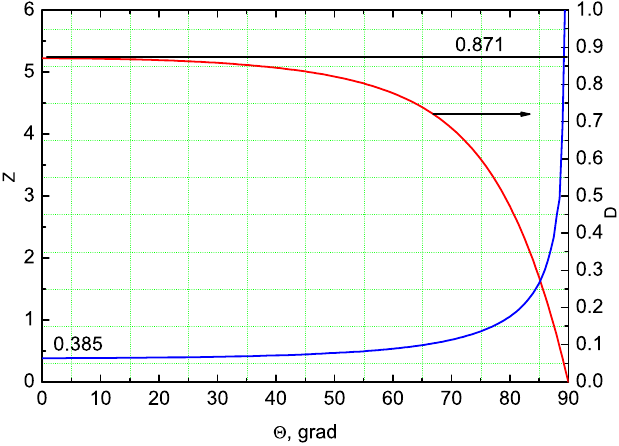}
\caption[]{1 (red curve) – Dependence of the transparency coefficient $D$ and the associated tunneling parameter $Z=\sqrt{(1/D)-1}$ (2, blue line) on the deviation of the electron incidence angle from the normal for the $Ta-Cu$ heterocontact, calculated from the tantalum side using the free-electron approximation.}
\label{Fig1}
\end{figure}

In heterocontacts, even in an ideal situation with no scatterers at the interface, a $\delta$-functional barrier arises due to the mismatch of the Fermi parameters of the contacting metals. The value of this barrier depends on the angle of incidence of the electrons relative to the normal at the interface. Figure \hyperref[Fig1]{1} (see Fig.1 in \cite{5}) shows the angular dependence of this barrier, determined by the transmission coefficient $D$ and the associated tunneling parameter $Z=\sqrt{(1/D)-1}$, calculated for a $Ta-Cu$ pair in the free-electron approximation (since a quantity such as the Fermi velocity cannot be precisely defined if the Fermi surface is not spherical \cite{11}). Therefore, if, when fitting the experimental curve for a given direction, all gaps have the same tunneling parameter $Z$, then they are represented for that direction. If, however, during fitting for a given gap the parameter $Z$ turns out to be close to 1, then the feature on the experimental curve associated with this gap is caused by the appearance of a tangential current component oriented parallel to the contact axis, due to current spreading.

\section{Experimental Methodology. Creation of point contacts and measurement of their characteristics.}\label{sec3}

When using massive electrodes to create point contacts, the insulating layer on their surface plays a crucial role. The area of the electrode contact zone is several orders of magnitude larger than the area of the resulting contact (short circuit). It is this insulating layer, called the support layer, that determines the mechanical and electrical stability of the resulting contact during measurements, as well as its electrical properties. Natural oxides are most commonly used as the support layer. For a wide range of metals, the use of metallographic etching modes to remove the defective surface layer formed during the process of shaping the electrode yields satisfactory results. In some cases, for certain rare-earth metals, etching requires the use of an alcohol-based electrolyte cooled to $-80^{\circ}$~C; other metals require two- or three-stage etching processes; and for some metals, standard recipes do not work, necessitating the empirical selection of the electrolyte composition and processing conditions.

When creating heterocontacts, copper or silver is typically used as the counterelectrode material. These materials are highly processable—the insulating layer on the electrode surface, consisting of natural oxides, possesses all the necessary properties. Furthermore, when in contact with transition metals or compounds based on them, the contribution of copper or silver to the resulting EPI spectrum is negligible.

When using bulk electrodes, the needle-anvil geometry was historically the first approach taken to create point contacts. This method has a number of inherent drawbacks. First, to break through the oxide layer and establish a short circuit, it is necessary to generate relatively high pressure at the contact point. For metals with a strong oxide layer on the electrode surface, this leads to the accumulation of crystal lattice defects in the vicinity of the short circuit and a high background level in the EPI spectra. Furthermore, in this geometry, frequent sample reheating and repeated electrochemical treatment are required to remove the resulting defect layer and restore the needle to its original shape (see, for example, Fig.8 in \cite{12}), which slows down the research process.

To address this shortcoming, a solution was employed that combines two types of deformation: compressive deformation and shear deformation \cite{13}. When using this method, a significantly lower electrode clamping force does not result in contact formation. A short circuit is formed due to the relative movement of the electrodes, which leads to the shearing off of the micro-irregularities that are always present on the surface. It was precisely the use of this method that allowed the author \cite{13} to obtain high-quality EPI spectra of aluminum for the first time, which had not been possible using the needle-anvil geometry. Note that in \cite{12} (Fig. 8), shear deformation is effectively employed, since when a long, thin needle is twisted into a spiral, the point of contact on the spiral section slips relative to the anvil.

Since not every point contact created possesses the required properties, from the experimenter’s perspective, the ideal technology would be one that allows for the creation and inspection of as many contacts as possible in a single experimental cycle, and the selection of the most suitable one from among dozens or even hundreds for more detailed investigation. It is important that the quality of subsequently created contacts not depend on their history. Sliding-contact technology best meets this requirement, and a special device was developed to take full advantage of its benefits \cite{14}.

Structurally, the device consists of an analytical unit and a removable shaft, connected by a ring-type electrical connector and mechanical coaxial connectors. The mechanical connectors transmit rotational motion through the shaft to the screw adjustments that control the movement of the upper electrode and the force with which it is pressed against the sample. The second electrode is rigidly mounted at the bottom of the analytical unit. The adjustable electrode is mounted on an $L$-shaped wire torsion damper; by twisting the damper’s arm, the electrodes can be moved apart or the contact force can be gradually increased. The relative positioning of the electrodes is such that, as the spring-loaded electrode moves, the contact area is constantly updated. The temperature is monitored by a semiconductor thermometer located at the intersection of the electrodes. The analytical unit is enclosed by a thin-walled copper tubular heater with a diameter of 18 mm. The removable shaft has a sliding coupler that allows the analytical unit to be moved between the warm and cryogenic regions of the cryostat. Rotational motion is transmitted through the shaft via thin-walled coaxial tubes driven by worm gears.

The device is equipped with an intermediate cryostat that allows measurements to be taken in the temperature range from 1.4 K to room temperature. Structurally, it consists of a tube with a diameter of 25 mm. Inside the lower part of the vacuum gap is a capillary restrictor with an inner diameter of 0.2 mm and a length of $\sim 1$ m. To reduce the cross-sectional area, a 0.1-mm-diameter wire is threaded through the capillary. The inlet of the capillary is protected by a glass filter, and the outlet is protected by a cotton ball.

\begin{figure*}[t!]
\centering
\includegraphics[width=1\linewidth]{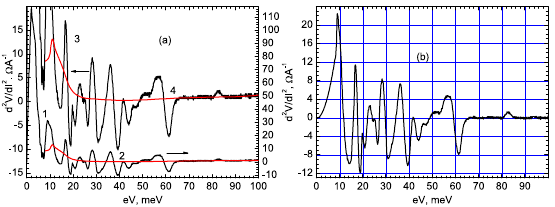}
\caption{
(a) 1 (black curve) – the second derivative of the $I-V$ curve for a point contact of $2H-NbSe_{2}–Cu$, $T=4.2 K$, $H=0$, with the contact axis oriented parallel to the layers. 2 (red curve) – the superconducting background. 3, 4 – the same plots on a larger scale.
(b) The second derivative of the $I-V$ curve after subtracting the superconducting background. The initial region from 0 to 8 mV was drawn by hand such that the geometric rule for sums is satisfied, allowing the integration of this curve to yield a zero background at the boundary of the EPI spectrum.
}
\label{Fig2}
\end{figure*}

Point contacts were formed between $Cu$ single crystals and $2H-NbSe_2$. The ratio of the resistance at room temperature to the residual resistance at low temperature was $\rho_{300}/\rho_{res}>1000$ for $Cu$ and $\rho_{300}/\rho_{res}\geq30$ for $2H-NbSe_2$. For measurements along the $NbSe_{2}$ layers, copper electrodes were cut on an electrical discharge machining (EDM) machine into bars $10\div15$ mm long and with a cross-sectional area of $1\times 1$ or $1.5\times 1.5\ \ m{{m}^{2}}$. For experiments perpendicular to the layers, the electrodes were cut into the shape of a pyramid with a base of $1\times 1 $ or $1.5\times 1.5 \ \ m{{m}^{2}}$ and a height of 3$\div$5 mm. The radius of curvature at the tip of the pyramid after final etching was approximately 0.1 mm or less. The defective layer formed after electrical discharge machining was removed by chemical or electrochemical etching. Surface quality was inspected using a metallographic microscope.

\begin{figure}[]
\includegraphics[width=8.5cm,angle=0]{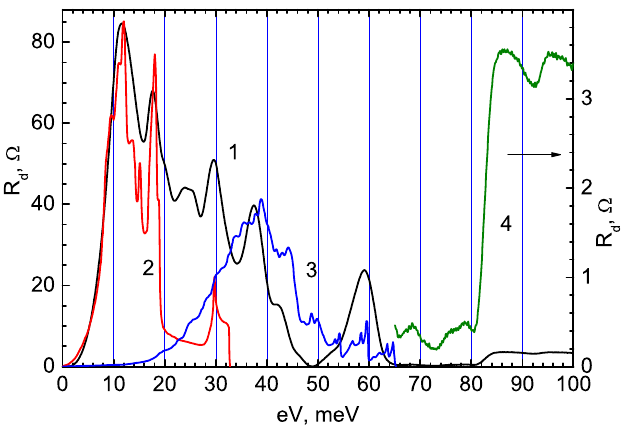}
\caption[]{
1 (black curve) – the relative change in the differential resistance of a point contact, proportional to the EPI function, including the region of two-phonon processes, obtained by integrating the curve in Fig. 2; 2 (red curve) – the phonon state density function from \cite{15}; 3 (blue curve) – a convolution obtained from the phonon state density function for two-phonon processes; 4 (olive curve) – a magnified section of curve 1 corresponding to a jump in differential resistance upon the transition of the superconducting region to a non-equilibrium state.
}
\label{Fig3}
\end{figure}

$NbSe_{2}$ electrodes measuring $\sim 3\times 5 \ \ m{{m}^{2}}$ were cut from a crystal blade $\sim 0.1$ mm thick and then bonded to the holder using silver paste. Since superconductivity is typically partially suppressed at the natural growth faces of a single crystal, the top layer was cleaved off immediately before assembly when measurements were taken perpendicular to the layers. For measurements in the $ab$ plane, the edge of a copper bar was brought into contact with the cut edge of the $NbSe_{2}$ plate. Similarly, by pressing the apex of the copper pyramid against the plane of the $NbSe_{2}$ crystal and sliding it parallel to the surface, contacts were created along the $c$-axis. By adjusting the contact pressure and shifting the electrodes, contacts with varying resistance were obtained for subsequent measurements.

Initial experiments demonstrated high lability of the second derivatives of $NbSe_{2}$ point contacts \cite{16}. Due to the lack of clear selection criteria, a significant part of the spectra presented there are highly variable. Nevertheless, Figs.4 and 5 show several reasonably reproducible second derivatives, the best of which is presented in this study.

\section{Experimental Results}\label{sec4}

\subsection{Determination of the EPI Spectrum}\label{subsec1}

$2H-NbSe_2$ is a highly anisotropic, easily cleavable layered superconductor with weak interactions between layers. The layers are formed by three-layer “sandwiches”: $Se$ layer, $Nb$ layer, $Se$ layer. In each of the monolayer structural units, the atoms form a close-packed triangular lattice. At low temperatures, the lattice parameters are $a\approx0.345$ nm and $c\approx1.254$ nm. In the unit cell of niobium diselenide $2H-NbSe_2$ along the $c$-axis, there are exactly 2 monolayers (triple $Se-Nb-Se$ layers) due to the specific nature of their mutual rotation and the translational symmetry of the crystal lattice \cite{15}. If we consider the first monolayer, the next (second) monolayer is rotated $180^{\circ}$ around the vertical axis relative to the first. Because of this $180^{\circ}$ rotation, the second monolayer is the mirror image of the first. In order for the structure to fully repeat itself in space (returning to the initial orientation of the first layer), the system must add a third layer, which will be identical to the first. Thus, the minimum repeating period (period of identity) along the crystallographic c-axis comprises two oppositely oriented monolayers. Together, they form a single unit cell. The interaction between $Nb$ and $Se$ atoms is covalent, while the interaction between neighboring selenium atoms located in different structural “sandwiches” is a weak van der Waals interaction.

Specific resistance at room temperature ${\rho _\parallel } \sim 2 \cdot {10^{ - 4}}\Omega  \cdot cm$; 
${\rho _ \bot } \sim {10^{ - 3}}\Omega  \cdot cm$ \cite{17}. In \cite{18}, for a high-quality single crystal with a resistance ratio ${\rho _\parallel }(300K)/{\rho _\parallel }(7.5K) = 43$, the mean free path of electrons at $7.5 \; K$ was determined to be: $l_{ab}\approx183$ nm. From this, taking into account the resistivity at room temperature, we can estimate the constant $\rho l  \sim 85 \cdot {10^{ - 12}}\Omega\cdot c{m^2}$ (for comparison, for ordinary metals this value is an order of magnitude smaller; for example, for tantalum $\rho l \sim 5.9 \cdot {10^{ - 12}}\Omega\cdot c{m^2}$, and for copper, $\rho l \sim 5.3 \cdot {10^{ - 12}}\Omega\cdot c{m^2}$. The coherence length along the layers for undisturbed $2H-NbSe_{2}$ is $\xi_{\parallel } \sim7.8$ nm; perpendicular to the layers, $\xi_{\bot }  \sim2.6$ nm, which coincides with the lattice period.
We will estimate the contact diameter when measuring along the layers using the Wexler interpolation formula \cite{19}:

\begin{equation}
\label{eq2}
d \simeq \frac{\rho }{{2R}} + \sqrt {{{\left( {\frac{\rho }{{2R}}} \right)}^2} + \frac{{16\rho l}}{{3\pi R}}}
\end{equation}

Using the parameters listed above and taking into account that, for our $2H-NbSe_{2}$ samples, the residual resistance is approximately $\sim 30$ times lower than at room temperature (${\rho _\parallel }\sim 6.6(6) \cdot {10^{ - 6}} \; \Omega  \cdot cm$), and the resistance $R_{0}^{N}\sim1080 \; \Omega$, we find that the diameter of the point contact is on the order of $d\sim6.33\; nm$.  As can be seen from the estimate, the contact diameter is determined solely by the Sharvin resistance, while the Maxwells contribution is negligibly small. This corresponds to a contact area $S\sim 31.47\; n{m}^{2}$. Given the layered structure of the material and the monolayer thickness $c\approx1.254$ nm, a contact of this area can be achieved using various combinations of three-layer “sandwiches,” ranging from a single element $\sim25\; nm$ long to to four three-layer structural elements $\sim6.27\; nm$ long. It is clear that combinations of elements of different lengths are more likely.

As noted in the introduction, a curve proportional to the EPI can be reconstructed from the superconducting additive in the excess current. To do this, the superconducting background must be subtracted from the difference between the second derivatives in the superconducting and normal states, and the resulting curve must then be integrated.

As the analysis shows, the dimensions of our contact are comparable to the coherence length. Using tantalum as an example, it was shown that the probability density for the wave functions of Andreev electrons is inversely proportional to the volume in which this conversion occurs. Thus, the magnitude of the superconducting nonlinearity is influenced here by two multiplicative factors: the high current density near the short-circuit point, which ensures a dense flow of non-equilibrium phonons, and the high probability density of the wave functions of Andreev electrons, on which these phonons scatter. Since the volume in which the nonlinearity is generated is small, there is no selection effect characteristic of tantalum contacts, and there is no need to use a correction curve when determining the EPI spectrum. Since the nonlinearity caused by backscattering processes is negligible compared to the superconducting nonlinearity, we use only the procedure of subtracting the superconducting background when reconstructing the curve proportional to the EPI function.

Fig.\hyperref[Fig2]{2(a)} shows the second derivative of the contact $I-V$ curve, with the axis oriented along the layers. The superconducting background is plotted here in such a way that, after it is subtracted, the geometric sum rule holds. The curve after subtraction of the background is shown in Fig.\hyperref[Fig2]{2(b)}. The initial section of the curve up to $\sim8\; mV$ in the energy gap region was plotted by eye to satisfy this rule. The superconducting background curve and the initial section of the curve complement each other to satisfy the sum rule; therefore, its shape does not affect the resulting EFV spectrum.

Other variations of the superconducting background, the initial segment, and the reconstructed electron-phonon interaction (EPI) function can be seen in Figs. 3 and 4 in \cite{20} and in Figs. 12 and 14 in \cite{21}. As can be seen from these figures, the positions of the phonon features and their relative intensities depend only slightly on variations in the superconducting background. 

The result of the integration (black curve 1) is shown in Fig.\hyperref[Fig3]{3}. The maximum energy of the change in relative differential resistance is $\sim64\; meV$, which is twice the maximum energy of the phonon density of states obtained in the theoretical study \cite{15}.

\begin{figure}[]
\includegraphics[width=8.5cm,angle=0]{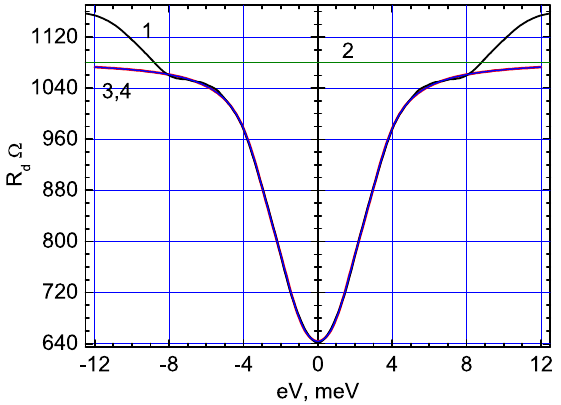}
\caption[]{
1 (black curve) – differential resistance of a point contact, obtained by integrating the second derivative of the $I-V$ curve shown in Fig.\hyperref[Fig2]{2(a)}; 2 (olive line) – estimate of the differential resistance of a point contact in the normal state at zero bias $R_0^N\sim1080 \ \ \Omega$; 3, 4 (blue and red curves) – calculations using the BTK formulas in the two- and three-gap approximations. Blue curve: $\Delta_{1}=1.31 \ \ meV$, $\Gamma_{1}=0.6 \ \ meV$, $k_{1}=0.82$; $\Delta_{2}=3.02 \ \ meV$, $\Gamma_{2}=0.3 \ \ meV$, $k_{2}=0.18$; $Z=0.2$. Scaling factor $S_{F} = 1.387$. Red curve: $\Delta_{1}=0.6 \ \ meV$, $k_{1}=0.388$; $\Delta_{2}=1.66 \ \ meV$, $k_{2}=0.404$; $\Delta_{3}=3.1 \ \ meV$, $k_{3}=0.208$; $k_{1}+k_{2}+k_{3}=1$; $\Gamma_{1}=\Gamma_{2}=0.01 \ \ meV$, $\Gamma_{3}=0.18 \ \ meV$, $Z=0.25$. Scaling factor $S_{F}\sim1$. 
}
\label{Fig4}
\end{figure}

S.~B.~Feodos'ev, a co-author of \cite{15} and this publication—who, unfortunately, has passed away—suggested that this might be related to two-phonon electron-phonon scattering processes. To support this hypothesis, he calculated the corresponding convolution of the phonon state densities, which perfectly confirmed his assumption (blue curve 3 in Fig.\hyperref[Fig3]{3}).

The step on curve 1 at energies above 80 meV is associated with the transition of the superconductor’s contact region to a non-equilibrium state, accompanied by a jump in the differential resistance. A similar jump for tantalum point contacts can be observed in Fig. 4 in \cite{5}, where the physics of such a transition is also discussed.

It should be noted that, for previously studied superconductors, two-phonon scattering processes on Andreev electrons were not observed, and their existence in $2H-NbSe_{2}$ appears to be related to the short coherence length in this compound.

\subsection{Calculation of superconducting energy gaps for the principal directions}\label{subsec2}

\subsubsection{\textbf{$ab$-plane}}

\begin{figure}[]
\includegraphics[width=8.5cm,angle=0]{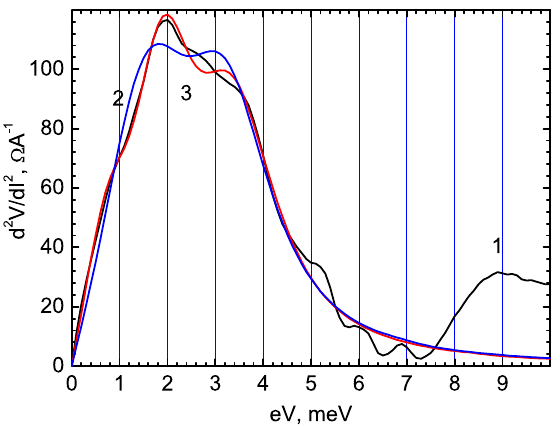}
\caption[]{
1 (black curve) – the second derivative of the $I-V$ curve (see Fig.\hyperref[Fig2]{2(a)}). 2 (blue curve) – calculation using the two-gap approximation. 3 (red curve) – calculation using the three-gap approximation.
}
\label{Fig5}
\end{figure}

\begin{figure*}[t!]
\centering
\includegraphics[width=1\linewidth]{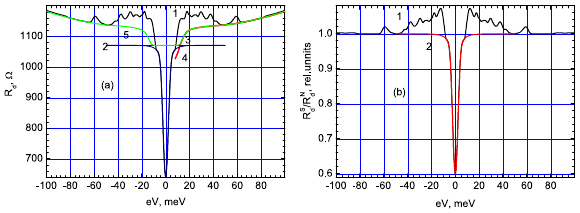}
\caption{
(a) 1 (black curve) – differential contact resistance, reconstructed from the second derivative (see curve 1 in Fig.\hyperref[Fig2]{2(a)}); 2 (navy curve) – calculation using BTK formulas in the three-gap approximation (the same as curve 4 in Fig.\hyperref[Fig4]{4} but in the range of $\pm50 \ \ mV$); 3 (blue segment) – an estimate of the differential contact resistance in the normal state at zero bias (the same as segment 2 in Fig.\hyperref[Fig4]{4}); 4 (red curve) – the result of integrating the superconducting background curve (the red curve in Fig.\hyperref[Fig2]{2(a)}); 5 (green curve) – the differential contact resistance in the normal state (formed by smoothly connecting segment 3 and curve 4 in the vicinity of their intersection).
(b) 1 (black curve) – normalized differential resistance of a point contact. 2 (red curve) – normalized three-gap theoretical curve.
}
\label{Fig6}
\end{figure*}

Fig.\hyperref[Fig4]{4} (curve 1) shows the differential resistance of a point contact along the layers, obtained by integrating curve 1 in Fig.\hyperref[Fig2]{2(a)}, as well as an estimate of the contact resistance in the normal state (segment of line 2). The calculation within the BTK model was performed using the two-gap and three-gap approximations (curves 3 and 4). As can be seen from the figure, the calculated curves practically coincide. The number of independent fitting parameters used when fitting curves in the two-slit variant 6: $\Delta_{1}$, $\Gamma_{1}$, $k_{1}$, $\Delta_{2}$, $\Gamma_{2}$, $Z$. In the three-gap version, there are 7: $\Delta_{1}$, $k_{1}$, $\Delta_{2}$, $k_{2}$, $\Delta_{3}$, $\Gamma_{3}$, $Z$. Not included here are $\Gamma_{1} = \Gamma_{2}\sim 0$, which are constants in the fitting process. In addition to their shape, the calculated curves are characterized by their amplitude (intensity), which is quantitatively described by the scaling factor $S_{F}$ mentioned in the introduction. Within the framework of the BTK model, the scaling factor cannot exceed 1. Based on this, the three-slot model should be preferred.

As noted in the introduction, a curve normalized to the normal state contains information about the scaling factor (the amplitude or intensity of the calculated curve relative to the experimental data). If only a superconducting experimental curve is available, there is a two-step procedure that allows one to estimate the normal resistance of a point contact at zero offset and the value of the scaling factor. To do this, the unnormalized curve is first fitted, and then, using the fitted parameters, a normalized theoretical curve is calculated with an arbitrary scaling factor. After that, the first curve is divided by the second. Through a series of successive approximations, by varying the scaling factors of the theoretical curve, the result of this division is made to approximate a line segment. Its position on the y-axis will represent an estimate of the normal resistance (curve 2 in Fig.\hyperref[Fig4]{4}). The accuracy of the SF determination is very high; a $1\%$ deviation transforms the result of dividing a line segment into an oscillating curve.

Note that for the second derivative, the deviation of the calculated curve from the experimental one is more pronounced than for the differential resistance. As can be seen from Fig.\hyperref[Fig5]{5}, the approximation using the two-gap approximation in the region of small offsets is unsatisfactory (curve 2), whereas the three-gap approximation works almost perfectly (curve 3).

Using the results obtained from determining the EPI spectrum, we can derive a number of additional relationships. First, in addition to the differential resistance across the entire energy range, Fig.\hyperref[Fig6]{6} shows an estimate of the differential resistance in the normal state within the same range. To determine this, we used the integral of the superconducting background curve in Fig.\hyperref[Fig2]{2(a)}, curve 4) and a segment of the normal resistance line at zero bias in Fig.\hyperref[Fig4]{4} (segment 2). By smoothly fitting them together in the vicinity of the intersection and symmetrically reflecting the resulting curve into the negative energy region, we obtain the desired curve 5.

\begin{figure}[]
\includegraphics[width=8.5cm,angle=0]{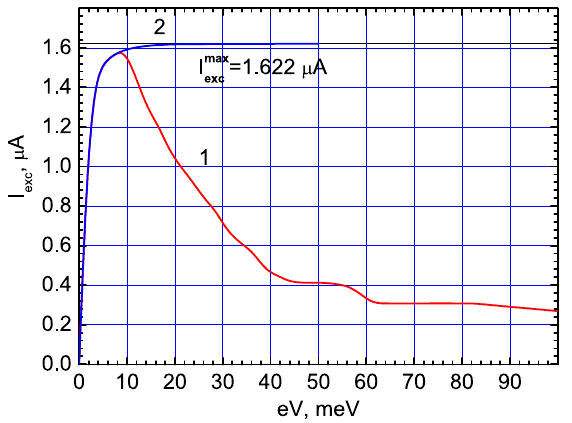}
\caption[]{
1 (red curve) – dependence of the difference between the $I-V$ curves in the superconducting and normal states (excess current) on the bias; 2 (blue curve) – the same quantity in the absence of EPI (the curves were obtained from the data shown in Fig.\hyperref[Fig.6]{6(a)}).
}
\label{Fig7}
\end{figure}

Using the differential resistance curves in the superconducting and normal states shown in Fig.\hyperref[Fig6]{6(a)}, the corresponding $I-V$ curves were reconstructed, and the dependence of the excess current on the contact bias was determined from their difference (Fig.\hyperref[Fig7]{7}). The break in curve 1 near 80 mV is caused by the transition of the near-contact region to a non-equilibrium state, leading to the appearance of a reverse current and the associated additional voltage \cite{5}. As already noted in the cited work, non-equilibrium features are observed only in samples with a sufficiently well-ordered crystal lattice.

\subsubsection{\textbf{$c$-direction}}

Results dedicated to the discovery of superconducting gaps in a direction perpendicular to the $2H-NbSe_{2}$ layers were published in a previous article \cite{5}, but have been supplemented to reflect new results related to measurements along the layers.

\begin{figure}[]
\includegraphics[width=8.5cm,angle=0]{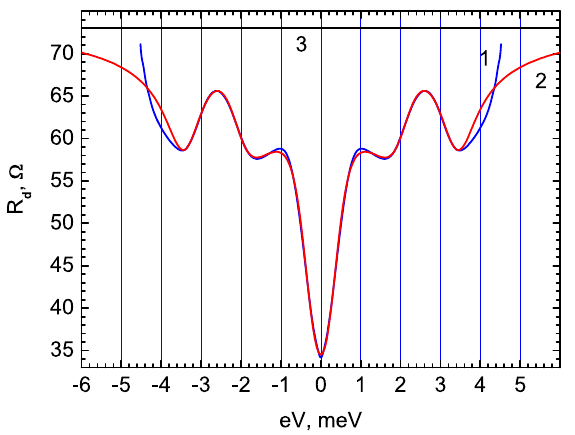}
\caption[]{
1 – Differential resistance of a $2H-NbSe_{2}–Cu$ point contact, $T = 1.65 \ K$, $H = 0$. 2 – Calculated curve: $\Delta_{1}=0.245 \ \ meV$; $\Delta_{2}=2.012 \ \ meV$; $\Delta_{3} = 3.46 \ \ meV$; $\Gamma_{1}=0.05 \ \ meV$; $\Gamma_{2}=0.104 \ \ meV$; $\Gamma_{3}=0.118 \ \ meV$; 
$Z_{1,2}=0.237$; $Z_{3}=0.985$; contribution from the gap $\Delta_{1}$, $k_{1}=0.71$; contribution from the gap $\Delta_{2}$, $k_{2}=0.2$; contribution from the gap $\Delta_{3}$, $k_{3}=0.09$. $k_{1}+k_{2}+k_{3}=1$
3 – calculated estimate of the differential resistance of a point contact in the normal state at zero bias $R_0^N\sim 73 \ \ \Omega$; scaling factor $S_{F}\sim 1.9$.
}
\label{Fig8}
\end{figure}

The estimated contact diameter $d \sim 8.5 \div 15 \ \ nm$, obtained in a previous publication, is many times greater than the coherence length ${\xi _c}(0) \simeq 2.6 \ \ nm$ in the direction perpendicular to the layers and coinciding with the lattice period. Thus, the fundamental condition for the BTK equation is violated here—the contact must be ballistic. Since this requirement is satisfied with respect to the pulse and energy relaxation lengths, there are no difficulties in fitting the shape of the theoretical curve to the experimental data here; however, this is not the case with respect to the scaling factor. Its value is close to 2. Furthermore, in a previous publication, it was only possible to fit the theoretical curve to the experiment up to 2.5 mV, provided that the same value of the tunneling parameter $Z$ was used for all gaps. If, however, this restriction is removed, the theoretical curve can be fitted to the experimental data practically up to 3.5 mV, near the breakpoint into a non-equilibrium state (Fig.\hyperref[Fig8]{8}). The extremely high value of the parameter $Z \sim 1$ for the third gap compared to the other gaps is due to the fact that this gap exists only in the ab direction and appears in the spectrum only as a result of current spreading, which leads to the emergence of its tangential component parallel to the ab plane. In hetero-contacts (see Fig.\hyperref[Fig1]{1}), a $\delta$-functional barrier exists at the interface, caused by the mismatch of the Fermi parameters of the contacting metals; its magnitude increases as the angle of incidence of electrons at the interface deviates from the vertical. As the angle approaches 90°, the transmission coefficient $D$ tends to 0, and the parameter $Z = \sqrt {(1/D) - 1}$ tends to infinity. In actual calculations, we obtained a parameter $Z\sim1$.

\begin{figure}[]
\includegraphics[width=8.5cm,angle=0]{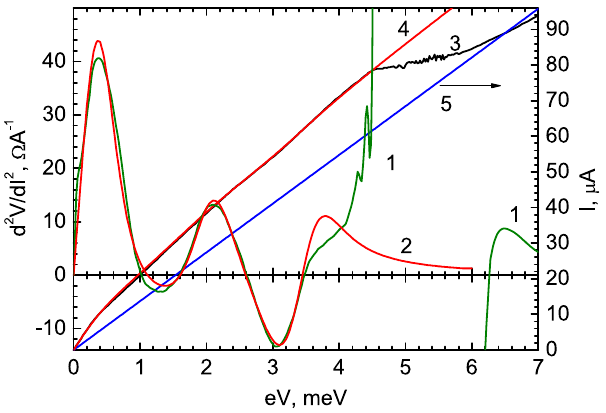}
\caption[]{
1 – second derivative of the $I-V$ curve (a fragment of the curve after breakdown is also shown), $T = 1.65 \ K$, $H = 0$. 2 – calculated curve. 3 – $I-V$ curve; 4 – calculated $I-V$ curve; 5 – estimated $I-V$ curve for the normal state, $R_0^N\sim 73 \ \ \Omega$.
}
\label{Fig9}
\end{figure}

Figure \hyperref[Fig9]{9} shows the $I-V$ curve for a point contact and its second derivative, their theoretical counterparts, as well as an estimated $I-V$ curve for the normal state. The break in the $I-V$ curve (transition to a new branch with a different slope) at energies above 4.5 meV is caused by the transition of the layer in contact with the copper counterelectrode to a non-equilibrium state with a partially suppressed gap and is discussed in more detail in a previous publication \cite{5}. 

\begin{figure}[]
\includegraphics[width=8.5cm,angle=0]{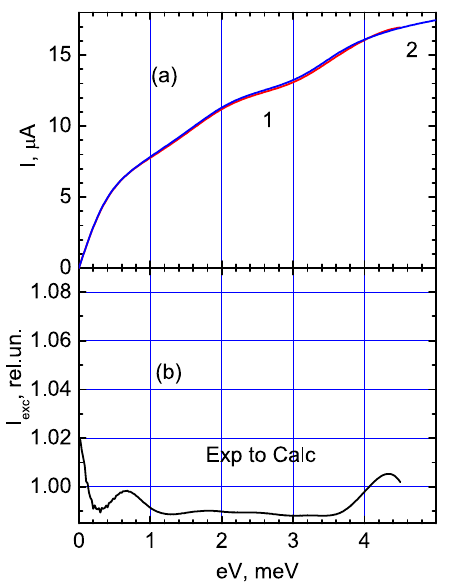}
\caption[]{
(a) – Excess current for the curves from the previous figure: (1 (red)) – experiment; (2 (blue)) – calculation. (b) – Relative discrepancy between the experiment and the calculation.
}
\label{Fig10}
\end{figure}

Finally, Fig.\hyperref[Fig10]{10} shows the difference between the $I-V$ curve (3) and the estimated $I-V$ curve (5) for the normal state (Fig.\hyperref[Fig9]{9}), referred to as the excess current. As can be seen from the figure, the deviation does not exceed $1 \div 2 \% $.

\section{Discussion of the Results}\label{sec5}
\begin{figure*}[t!]
\centering
\includegraphics[width=0.8\linewidth]{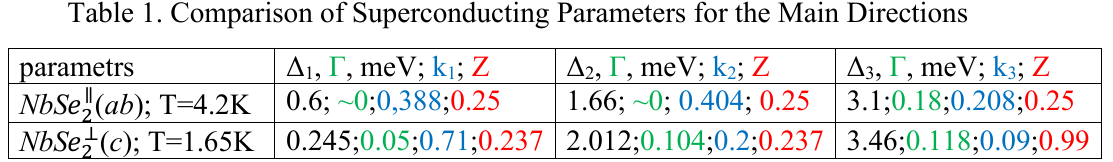}
\label{Table1}
\end{figure*}

In point-contact spectroscopy, it is possible to assess the perfection of the crystal lattice in the region where non-equilibrium phonons are generated based on the shape of the second derivative of the $I-V$ curve. Lattice distortions lead to the blurring of phonon features, up to their complete disappearance, especially in small contacts. If the nonlinearity in the contact is caused by backscattering processes, then information about the lattice perfection comes from the region of current concentration near the short-circuit point. The BTK theory is based on the assumption of ballistic transport in contacts. At the same time, applying the equations of this theory to determine superconducting parameters in contacts that are “dirty” from the perspective of PC spectroscopy generally yields excellent results. In a review by Naidyuk and Glooss (see Fig.14 in \cite{22}) that applying this theory to metals with a large coherence length in such contacts is justified, since the metal’s crystal lattice remains undistorted over most of the volume where the conversion of Andreev electrons into Cooper pairs occurs and where the gap nonlinearity is formed. 
Moreover, the first experiments conducted by Blonder and Tinkham \cite{12}—which quantitatively confirmed her conclusions—were performed on $Nb$-based point contacts that do not possess an extremely long coherence length. It is extremely difficult to obtain an EPI spectrum using such contacts; we are not aware of any studies in which high-quality point-contact EPI spectra of niobium have been obtained. 

We expended a great deal of effort trying to solve this problem, but the result obtained was rather mediocre \cite{8}. Nevertheless, in the superconducting state, when the volume in which the EPI nonlinearity is formed does not coincide with the volume of backscattering processes, it becomes possible to recover features in these contacts that are proportional to the EPI function. This experimentally confirms the hypothesis put forward in \cite{22} regarding the decisive role of the undisturbed metal in the formation of the gap feature.

In the case of $NbSe_{2}$, due to the short coherence length, the volume in which the gap feature is formed—that is, the volume in which the conversion of Andreev electrons into Cooper pairs occurs—coincides with the volume in which the nonlinearity caused by EPI is formed. This circumstance leads to an unambiguous conclusion: for the information about the gap parameters in $NbSe_{2}$ to correspond to an unperturbed crystal lattice, it is necessary and sufficient that the point contact under study exhibit sharp phonon features.

Unlike previously studied superconductors, the features in the second derivative of the current-voltage curve of an $NbSe_{2}$ point contact extend all the way up to twice the energy of the phonon density of states. Moreover, the amplitude of these features is of the same order of magnitude across the entire energy range. S.B.~Feodosyev’s hypothesis regarding the two-phonon nature of the features at energies above the phonon density of states energy threshold, supported by the two-phonon convolution he calculated, perfectly describes the experimental curve across the entire energy range. This allows us, first, to hypothesize that the features beyond the energy density of the phonon states are associated with phonon quasi-particle excitations, and second, since not all experimental curves exhibit this behavior, to use the presence of such features in the spectrum as a criterion for the quality of point contacts. In other words, the absence of such features may be related to crystal lattice perturbations and can serve as a selection criterion.

Higher temperatures and higher point-contact resistance contributed to the absence, in the phonon frequency range, of nonlinearities associated with non-equilibrium processes that interfere with the study of electron-phonon interactions using perpendicularly oriented contacts.

Table \hyperref[Table1]{1} lists the superconducting gaps parameters for point contacts in the two main crystallographic orientations.

If $\Delta(1.65 K) \approx \Delta(0)$, then $\Delta(4.2 K) \approx 0.9\Delta(0)$. Thus, for the $ab$-plane at a comparable temperature, $\Delta_{1} \approx 0.67 meV$; $\Delta_{2} \approx 1.84 meV$; and $\Delta_{3} \approx 3.44 meV$.

In heterocontacts, a potential barrier arises in the absence of any scattering media between the metals. The barrier height is minimal when electrons strike the interface vertically. As the angle of incidence deviates from the vertical, the barrier height increases, reaching a maximum at tangential incidence (Fig.\hyperref[Fig1]{1}). As shown in the table\hyperref[Table1]{1}, the tunneling parameter $Z$, which is related to this barrier, is nearly identical for both directions, with the exception of the gap $\Delta_{3}$ for the $c$-direction. Current spreading in a contact whose axis is oriented in the $c$-direction leads to the appearance of a tangential current component directed parallel to the $ab$-plane. The correlation between the value of the $Z$ parameter and the direction of the current allows us to conclude that the gap with an energy of $3.46 / meV$ is associated with the $ab$-plane.

It is evident that the spreading current, which causes features related to the $ab$-direction to appear on the curve in the $c$-direction, applies to all gaps, not only to gap $\Delta_{3}$, which is considered in this calculation. It is very difficult to perform calculations using an approximation involving more than three gaps, but their presence on the experimental curve can be inferred from indirect evidence.

To ensure a correct analysis of our experimental results, we present data for the $c$-direction obtained on flat tunneling structures \cite{23} using a scanning tunneling microscope with a Pt-Ir tip \cite{24} and using a scanning tunneling microscope with a superconducting $Nb$ tip \cite{25}. The data obtained on planar structures and using a microscope with a Pt-Ir tip indicate the presence of two gaps: $\Delta_{1} \sim 0.7\div0.9 / meV$ and $\Delta_{2} \approx 1.2 \div 1.4 / meV$. At the same time, such structures are particularly flat and, for a number of reasons (Tunnel current in STM and planar contacts is extremely sensitive to the properties of the first few atomic layers. Oxidation, adsorbates, or mechanical strain alter the density of states on the surface), exhibit lower values of the superconducting gaps.

Data obtained using a scanning tunneling microscope \cite{25} with a superconducting niobium tip show significantly less suppression of the $\Delta_{2}$ gap. The maximum experimental value of the gap distribution edge is approximately 1.55 meV (within a range up to 1.7 meV). Maximum calculated value (SCDFT): In the initial (unscaled) theoretical calculation, the distribution of the gap values across the entire Brillouin zone has a main peak at 2.3 meV, and the average value is \(\bar{\Delta} = 1.85\text{ meV}\). In the simulated local STM spectra for the most active layer (\(L_{2}\)), an additional shoulder-like feature is also observed in the region of \(\approx 3\) meV.

Let’s return to our experiments (Fig.\hyperref[Fig8]{8}). Since the $NbSe_{2}$ crystal cleaves along weak van der Waals planes, the copper electrode comes into contact with the $Se$ layer. Due to the proximity effect upon contact with copper, superconductivity in this layer is partially suppressed, which leads to an anomalously small value of $\Delta_{1}$ for the $c$-direction. Since information about the gap comes from a distance equal to the coherence length $\xi\simeq 2.5 / nm$, it includes 2 monolayers, i.e., 4 layers of $Se$. If the first layer has a gap suppressed by the proximity effect, then the subsequent layers—at least in the second monolayer—must have a maximum equilibrium value of $\sim0.7\div0.9 / meV$. Furthermore, the gap from the $ab$-plane, $\Delta_{1} \sim 0.67 / meV$, which arises due to current spreading, falls within this same interval. We approximate this entire set of gaps, with their individual contributions, using a single gap $\Delta_{1} \simeq 0.245 / meV$ and a broadening parameter $\Gamma_{1} \sim 0.05 / meV$, i.e., $\sim20\%$ of the gap value. Despite careful fitting, we were unable to perfectly align the theoretical and experimental curves in shape, as shown in Fig.\hyperref[Fig8]{4}. As in the case of the $ab$-plane (Fig.\hyperref[Fig5]{5}), the most noticeable deviation of the fit from the experiment for the $c$-direction can be seen in the second derivative (Fig.\hyperref[Fig9]{9}).

As for the contribution to the conductivity from the first gap, which is $k_{1}\sim0.7$, it is in perfect agreement with the contact geometry perpendicular to the layers (along the \(c\) axis): the selenide layers (\(Se-Nb-Se\)) shield the inner niobium layers. When current is injected perpendicular to the plane, electrons first tunnel into or transition to the Se-derived zone, which is elongated along the \(c\) axis. This zone has a larger projection of the Fermi velocity \(v_{Fz}\) along the contact axis than the quasi-two-dimensional $Nb$ cylinders. Consequently, the geometric conductivity factor of ballistic electrons ensures that the small gap dominates the integral current of the point contact.

Let us now turn to the larger gap $\Delta_{2}$. The maximum value obtained in the tunneling experiment \cite{25} is $\sim1.55 / meV$ (within a range up to 1.7 meV), which is significantly smaller than the $\Delta_{2}$ of $\sim 2.012 / meV$ obtained in our experiment. As noted above, in the unscaled theoretical calculation, the distribution of gap values across the entire Brillouin zone has a main peak at 2.3 meV. The 2.3 meV peak itself is physically associated with a “hot spot” in momentum space on the ML line of the Brillouin zone. First-principles calculations (SCDFT) overestimate the critical temperature (\(T_{c}\)) of the material. The model gives \(T_c = 10.2\text{ K}\) instead of the experimental \(7.2\text{ K}\). The reason for the overestimation of \(T_c\) is the anharmonic corrections to the dynamics of the crystal lattice. These effects increase significantly near the transition to the charge density wave (CDW) phase. In point-contact experiments, uniaxial compression occurs, increasing the local pressure in the contact region. CDWs compete strongly with superconductivity in the hollow-centered (HC) regions of the crystal, cutting off the density of states at the Fermi level (\(N_{F}\)). Pressure-induced suppression of the CDW eliminates this competition. As a result, the local \(N_{F}\) is restored, which locally strengthens the electron-phonon interaction, increasing both the local critical temperature (\(T_{c}\)) and the superconducting gap (\(\Delta \)) in the contact region.

In other words, compression along the c axis reduces the interlayer spacing. This induces quantum melting or a strong CDW phase transition. Local lattice deformation weakens the CDW correlations and increases $T_{c}$ and $\Delta$ in the contact region. Obviously, the extent of the increase in $T_{c}$ and $\Delta$ depends on pressure and affects the first monolayer to the greatest extent. We approximate this set of gaps, with their individual contributions, as well as the temperature-corrected gap $\Delta_{2} \sim 1.84 \ meV$ in the ab plane—which arises due to current spreading—with a single gap $\Delta_{2} \sim 2.012 \ meV$  and a broadening parameter $\Gamma_{2}\sim 0.104 / meV$. As for the scaling factor $S_{F} \sim 1.9$ obtained with these approximations, one possible reason for its overestimated value is the same as when attempting to describe the experimental curve in the $ab$-plane (Fig.\hyperref[Fig4]{4},Fig.\hyperref[Fig5]{5}) using the two-gaps approximation.

In summary, the conclusion drawn in \cite{26} is confirmed: a large gap appears along the $c$-axis, whereas a smaller gap is observed in the $ab$-plane.

In our study of the anisotropy of superconducting gaps in $NbSe_{2}$, we took full advantage not only of the qualitative agreement between the BTK calculation and the experiment (the similarity in the shape of the curves), but also of the quantitative agreement (taking the scaling factor into account).

\section{Conclusion}\label{sec6}

\begin{enumerate}
\item The spectrum of electron-phonon interaction in $2H-NbSe_{2}$ was obtained; it was found that the short coherence length results in similar intensities for single- and two-phonon scattering processes, and criteria were established indicating that the crystal lattice is undisturbed in the region of phonon detection and superconducting gaps.

\item It was found that the superconducting energy gaps are anisotropic; specifically, there are three gaps in the $ab$ plane, while there are only two in the $c$-direction. The possibility of detecting a third gap in the contact differential resistance curve in the $c$-direction is due to the presence of a tangential current component arising during spreading, as evidenced by the large value of the tunneling parameter $Z$ for this gap.

\item It has been established that for the $ab$ direction, when the contact size and coherence length are comparable, the BTK equations remain applicable both in terms of the shape of the calculated curves and in terms of their intensity (scaling factor). For the $c$ direction, for which the coherence length becomes significantly smaller than the contact size, the question of the intensity of the experimental curve remains open.

\end{enumerate}

This work was supported by the National Academy of Sciences of Ukraine as part of project F19-5.

In conclusion, we wish to thank A.~V.~Terekhov for numerous tips, comments and additions made during discussion of the paper. 

\bibliography{NbSe2.bbl}

\end{document}